\documentstyle[12pt,a4,epsfig]{article}

\newcommand{\beq}[1]{\begin{equation}\label{#1}}
\newcommand{\eeq}{\end{equation}}
\newcommand{\beqar}[1]{\begin{eqnarray}\label{#1}}
\newcommand{\eeqar}{\end{eqnarray}}

\newcommand{\nn}{\nonumber}

\def\l{{\bf l}}
\def\k{{\bf k}}
\def\q{{\bf q}}
\def\r{{\bf r}}

\def\e{{\bf e}}

\newcommand{\al}{\alpha}
\newcommand{\be}{\beta}

\newcommand{\ga}{\gamma}
\newcommand{\de}{\delta}

\newcommand{\la}{\lambda}

\newcommand{\si}{\sigma}
\newcommand{\Ga}{\Gamma}

\newcommand{\La}{\Lambda}

\newcommand{\as}{\alpha_S}

\begin{document}
\vspace*{-2cm}
\begin{flushright}
NTZ 20/98 \\

\end{flushright}
\vspace{2cm}
\begin{center}
{\LARGE \bf

Polarization  in diffractive electroproduction \\
of light vector mesons\footnote{Supported by
German Bundesministerium f\"ur Bildung, Wissenschaft, Forschung und
Technologie, grant No. 05 7LP91 P0}}\\[2mm]
\vspace{1cm}
D.Yu.~Ivanov$^{\dagger \$}$ and
R. Kirschner$^{\dagger }$

\vspace{1cm}

$^\dagger$Naturwissenschaftlich-Theoretisches Zentrum  \\
und Institut f\"ur Theoretische Physik, Universit\"at Leipzig
\\

Augustusplatz 10, D-04109 Leipzig, Germany
\\
\vspace{2em}
$^{\$}$
Institut of Mathematics, 630090 Novosibirsk, Russia \\

\end{center}

\vspace{1cm}
\noindent{\bf Abstract:}
 We study in perturbative QCD 
the helicity ampltiudes of the process $\gamma^{*} p
\rightarrow  \rho p $ at large virtualities $Q$ of the photon
$\gamma^{*}$. We estimate all spin flip amplitudes taking into 
account an important effect of the  
scale behaviour of the gluon density. 
The transition of a transverse
virtual photon to a longitudinal vector meson is not small at typical
HERA conditions. This helicity non-conserving amplitude leads by
interference to a measurable effect in the distribution of the angle
between the electron scattering and the meson production planes.

\vspace*{\fill}
\eject
\newpage

\section{Introduction}

The process of diffractive $\rho$ meson electroproduction
\beq{1}
\ga^* p \to \rho p
\eeq
is now under intensive experimental study at the $ep$ collider HERA
in a new energy range as compared to the previous fixed target
experiments.
The word diffractive means that the energy of  $\ga^* p$ collision $W$
is much larger than the virtuality
of the photon $Q$. Therefore this process is mediated by the pomeron exchange.
On the other hand the photon virtuality is large in comparison
with the typical hadron scale and we deal here with hard diffraction.

The diagram of this process and our notations are given at Fig.1
\begin{figure}[htb]
\begin{center}
\epsfig{file=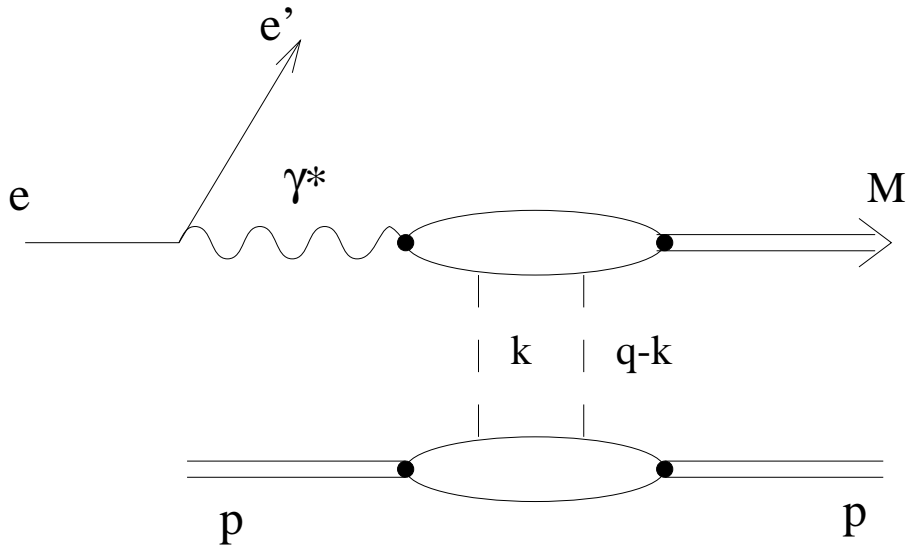,width=12cm}
\end{center}
\vspace*{0.5cm}
\caption{\label{fig-1}
Diffractive electroproduction of vector meson on a proton target.
}
\end{figure}

Theoretically this process was considered both in nonperturbative
and perturbative approaches in many papers. 
The references can be found in the review \cite{c}.
It turns out that the
perturbative models where pomeron is represented by the
hard two--gluon exchange are able to reproduce the
main features of the HERA data for hard diffraction. In these
models the amplitude is proportional to the gluon distribution
in the proton
\beq{2}
M\propto xG(x,\tilde Q^2) \ ,
\eeq
where $x=\frac{Q^2}{W^2}$, $\tilde Q$
is the typical hard scale of the process.

The aim of this paper is to consider in the framework  of
perturbative QCD the polarization effects in
diffractive electroproduction of vector meson
at large $Q^2$. Experimentally, information about the polarization
state of produced meson is extracted from the angular
distributions of the meson decay products ($\pi^+\pi^-$ for $\rho$).
The present day results of such investigations at HERA are
consistent with the $s$- channel helicity conservation (SCHC),
which means that the produced meson retains the helicity of incoming
virtual photon. The expected increase of the  HERA luminocity gives us a
hope that we will have  in the near future
much more precise experimental information
about helicity properties of the $\ga^* M$ transition.
Such an information will give us  valuable insight into the underlying
dynamics.

We shall discuss experiments with unpolarized protons,
therefore the proton can be formally considered as a spinless target and
we shall indicate further only the polarization states of the
virtual photon and the produced meson.

Under the asumption of SCHC there are only two independent helicity
amplitudes $M_{\ga^*_L\to \rho_L}$ and $M_{\ga^*_T\to \rho_T}$, where $L(T)$
denotes longitudinal(transverse) polarized state.

The longitudinal
amplitude dominates at large $Q^2$. It has been considered
in the papers \cite{r,br,fks,kolia,mrt}. A QCD factorization theorem for this
amplitude was proven in \cite{theorem}. The situation for the transverse
amplitude is more complicated. Formal power counting gives
$\sim \frac{m_\rho}{Q}$ suppression in comparison with
the longitudinal amplitude. However, this suppression factor is too small
for typical values of $Q^2\sim 10 \mbox{ GeV}^2$ for HERA
electroproduction experiments and does not agree with the measured
ratio of $\frac{\sigma_T}{\sigma_L}$. For the transverse amplitude
the integral over
the longitudinal fraction $z$ of the quark momentum is logarithmically
divergent in the end points of the integration region. This means
that essential transverse distancies which scale like
$\frac{1}{Q\sqrt{z(1-z)}}$ can be large even for large
$Q^2$ and that the transverse amplitude can receive large
or even dominant contribution from the nonperturbative region.
As was noticed in \cite{mrt}, if the nonperturbative contribution would be
dominant we would expect for $\sigma_T$ features
similar to the ones observed in photoproduction experiments as:
a "softer" as compared to $\si_L$ energy behaviour $\sim W^{0.2}$
and a larger slope $b\sim 9-10 \mbox{ GeV}^{-2}$.
However these expectations are not supported by the data.

In \cite{mrt} it was assumed that the scale behaviour
of gluon distribution
$G(x,Q^2)\sim \left(\frac{Q^2}{Q^2_0}\right)^\ga$,
where $\ga$ is the anomalous dimension of gluon
density, plays a very important role in the physics
underlying the transverse amplitude. Taking into
account this dependence it can be seen that
the typical transverse distances are $\sim \frac{1}{Q\sqrt{\ga}}$
and can be smaller than $\frac{1}{\La_{QCD}}$ at large $Q$
even for small $\ga$. Therefore perturbative QCD
can be applicable to the transverse amplitude. The
estimate derived in \cite{mrt} under the assumption of
constant anomalous dimension
\beq{3}
\frac{\si_L}{\si_T}=\frac{Q^2}{M_\rho^2}\left(
\frac{\ga}{1+\ga}
\right)^2
\eeq
shows the role of the scaling violation of the
gluon density and agrees qualitatively with the data.

Under the assumption of natural parity exchange in the $t-$
channel there are five independent helicity amplitudes.
These are the two  helicity conserving amplitudes
\beq{4}
M_{(0,0)}=M_{\ga^*_0\to \rho_0} \ , M_{(+1,+1)}=M_{\ga^*_{+1}\to \rho_{+1}} \
(M_{(-1,-1)}=M_{(+1,+1)}) \ ;
\eeq
and the amplitudes violating SCHC:
two single spin--flip amplitudes
\beq{5}
M_{(+1,0)}=M_{\ga^*_{+1}\to \rho_0} \ ,
(M_{(-1,0)}=-M_{(+1,0)}) \ ,
\eeq
\beq{6}
M_{(0,+1)}=M_{\ga^*_0\to \rho_{+1}} \ ,
(M_{(0,-1)}=-M_{(0,+1)}) \ ;
\eeq
and one helicity double--flip amplitude
\beq{7}
M_{(+1,-1)}=M_{\ga^*_{+1}\to \rho_{-1}} \ ,
(M_{(-1,+1)}=M_{(+1,-1)}) \ .
\eeq

As it will be shown below, similar to the non--flip amplitudes
(\ref{4}), the single spin--flip
amplitudes can be expressed through the gluon density in the proton.
In the leading order of $1/Q$
expansion the  double spin--flip amplitude does not receive
a logariphmic contribution from the integration over the $t-$
channel gluon momenta, see Fig. 1. Therefore in the first term of 
the $1/Q$ expansion of the double spin--flip
amplitude the  large factor $x G(x, \tilde Q^2)$ is absent.

We will show that
the largest amplitude violating SCHC is $M_{(+1,0)}$.
The other amplitudes
violating
SCHC can be neglected in a first approximation.
Our result derived in the approximation of constant gluon anomalous
dimension $\ga$ reads
\beq{8}
\be = \frac{M_{(+1,0)}}{M_{(0,0)}} =
\frac{ \sqrt{|t|} }{ \sqrt{2}Q\ga } \ .
\eeq
The observed $t-$ dependence for $\rho$ meson electroproduction
is $\frac{d\si}{dt}\sim e^{-bt} $, with  the slope
$b=5\dots 6\mbox{ GeV}^{-2}$.
At typical values of $t\sim 1/b$ this amplitude is not too small
and, as will be shown below, leads to a sizable interference effect.

In the perturbative QCD approach
the effect of the scale behaviour of the  gluon density manifests
itself qualitatively in the similar way in  both transverse
amplitudes: the helicity non--flip  $M_{(+1,+1)}$ and
the helicity single--flip
amplitude $M_{(+1,0)}$.
Therefore the mesurement of
$M_{(+1,0)}$ at HERA would give us an important check of whether
perturbative QCD  describes correctly
the  physics  underlying the amplitudes of vector meson electroproduction
initiated by a transverse photon.

The  present work is based on the experience gained in a 
previous study of the diffractive vector meson production \cite{gi}.

The paper is organized as follows. In section 2 we present the main steps
of our  calculation method.  The results for helicity amplitudes
and the discussion
of the underlying physical effects are given in section 3.
The influence of the helicity--flip amplitudes  on the
vector meson product distributions is discussed  in the section 4.
Our conclusions are summarized in section 5.

\section{ Impact
parameter representation and the meson wave functions}
\setcounter{equation}{0}

The amplitude of the diffractive process
$\ga^* p\to \rho p$ can be represented
as the integral over the transverse momenta of gluons in the $t-$
channel (impact representation)
\beq{9}
M_{\ga^* p\to \rho p}= is_{\ga^* p}\int
\frac{d^2 k}{\k^2(\q-\k)^2} J_{\ga^*\rho}J_{p} \ .
\eeq
Here $\q$  is the momentum transfer which is  transverse
 with  high accuracy, $q^2=t=-\q^2$.
Throughout the paper all vectors, if it is not mentioned separately,
are  two-dimensional
vectors in the transverse space.
The accuracy of the representation (\ref{9}) is expected to be  $\sim
Q^2/s_{\ga^* p}$.

The space--time  picture of the process in diffractive
(high energy) region is the following. The virtual photon
fluctuates into the $q\bar q$ pair long time before and the $q\bar q$
pair converts into the vector meson long time after the
interaction with the proton.
Therefore
it is possible to represent the photon impact factor
 as the convolution
of the impact factor for the $q\bar q$ dipole scattering
with the light cone wave functions of the incoming
virtual photon and the  outgoing vector meson (Fig.2)
\beq{10}
J_{\ga^*\to \rho}(\k,\q)=\int \frac{d^2 l_1dz_1}{16\pi^3}
\frac{d^2 l_2dz_2}{16\pi^3}
\Psi_{\ga^*} (\l_1,z_1)
\Phi^{dipole}
(\l_1,\l_2,z_1,z_2,\k,\q)\Psi_{\rho}^* (\l_2,z_2) \ ,
\eeq

where
\beqar{11}
& \Phi^{dipole}(\l,\l_1,z,z_1\k,\q)=16\pi^3\frac{\as \de ^{ab}}{N}
\de (z-z_1)
 \left[\de (\l - \l_1 - \q z)  +   \right. & \nn \\
&  \left. \de (\l-\l_1+\q \bar z) -
\de (\l - \l_1 + \k -\q z) - \de (\l - \l_1 -\k + \q \bar z)\right]
\ . &
\eeqar
$N=3$ is the number of colors, $(\de^{ab})^2=N^2-1$.
$\as$ is the strong coupling
constant.

\begin{figure}[htb]
\begin{center}
\epsfig{file=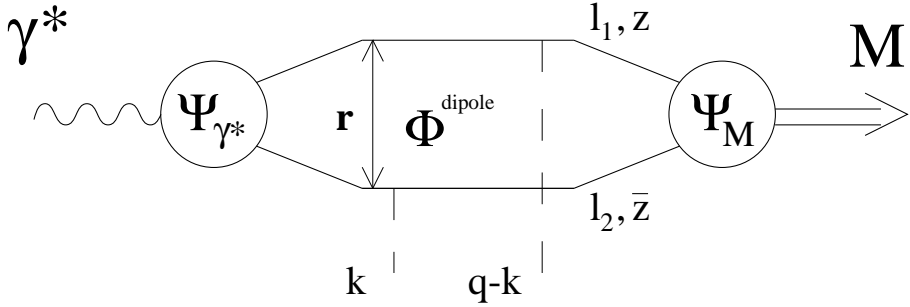,width=12cm}
\end{center}
\vspace*{0.5cm}
\caption{\label{fig-2}
The $\ga^* \to M$ impact factor. Only one contribution of the interaction of
the $q\bar q$ dipole with the two exchanged gluons (dashed lines)
is shown.
}
\end{figure}

The light cone wave function of the  photon
\beq{12}
\Psi_{\ga^*} (\l,z)=-e_q\sqrt{z\bar z}
\frac{\bar u\hat e v}{\l^2+Q^2z\bar z}
\eeq
describes the  probability amplitude for the  splitting of the
photon  into the $q\bar q$ pair with electric charge $e_q$,
($e_q=\frac{2}{3}e$ for the $u$ quark).
The quark carries the transverse momentum  $\l$ relative to
the photon momentum and  the fraction $z$ of
the photon longitudinal momentum
(the fraction for antiquark is $\bar z=1-z$).
It should be noted that the vector $\l_2$ in eq. (\ref{10}) is
transverse relative
to the momentum of the outgoing vector meson which
received non--zero momentum transver $\q$.

The polarization state of the photon is described
by the vector  $\e$.
We choose the following convention for
the polarization four vectors describing
a transversly polarized vector meson and a transverse photon
\beq{13}
\e_{\pm }=\frac{\mp 1}{\sqrt{2}}(0,1,\pm i,0) \ ,
\eeq
where the $x$ axis is choosen in the direction of the momentum transfer
\beq{14}
\q=q\cdot (0,1,0,0) \ .
\eeq
It can be seen that the difference in the polarization vectors
of the vector meson from (\ref{13})  related to the
non--zero value of momentum transfer can be neglected.
The polarization vector of the longitudinally polarized virtual photon is
\beq{15}
\e_0=\frac{1}{Q}(p_1,0,0,\sqrt{p_1^2-Q^2}) \ ,
\eeq
where $p_1$ is the value of the longitudinal momentum of the photon.
And the
similar convention is adopted for the  polarization vector of
the longitudinaly polarized vector meson.

The perturbative expression (\ref{12}) for the
splitting of a photon into  the light $\bar q q$ pair
has the important property that the total helicity of the produced
massless quark--antiquark pair is zero irrespective of the polarization
state of the photon: \\
a) for longitudinally polarization
\beq{16}
\sqrt{z\bar z}
\bar u_\la \hat e_0 v_{\la^\prime }=2Qz\bar z
\de_{\la,-\la^\prime} \ ,
\eeq
b) for transverse polaritation
\beq{17}
\sqrt{z\bar z}
\bar u_\la \hat e_\pm v_{\la^\prime }=\de_{\la, -\la^\prime}
\{(1-2z)\mp \la \} (\e_{\pm} \l ) \ ,
\eeq
The quark helicities $\pm \frac{1}{2}$
are represented  in the above
equations by $\la =\pm 1$.

The splitting of the transverse photon into the
quark pair with the total helicity $\pm 1$ is proportional
to the current quark mass. Since  light
quarks have very small current masses we will neglect here
this splitting. Of course it becomes important in the case of a
heavy flavour production.

The physics of the spin flip transitions which is
the main subject of this paper looks more transparent
in the space of  impact parameters of the $q\bar q$ pair.
The helicities of quarks coincide in this case with the
projection of the quark spins onto the $z$ axes. Since the total helicity
of the quark pair is zero, the projection of the orbital momentum of the
incoming  pair
onto the $z$ axis should coincide with the helicity of the incoming
virtual photon. Another important property of
perturbative QCD is that the interaction of $t-$ channel gluons with
the pair does not change with  high accuracy the helicity
states of the quarks. This interaction does not change also the
impact parameters of the pair even if the  momentum
transfer is not zero.
Therefore the helicity state of the produced meson should coincide with
the projection of the  orbital momentum of the outgoing quark pair onto
the $z$ axis.
We can conclude that in the frame of perturbative
QCD the only posibility to have the change of the helicity
state during the diffractive  $\ga^* \rho$ transition is to
change the $z$ projection of the  angular momentum of the
$q\bar q$ pair in the interaction
with the proton.

To make this discussion more quantitative let us transform
eqs. (\ref{10},\ref{11},\\
 \ref{12}) into the space of impact parameters.
By Fourier transformation we obtain the photon (meson)
wave function in the  representation of impact parameters
\beq{18}
\Psi (\r,z)=\int \frac{d^2\l}{2\pi}\Psi (\l,z)\cdot
e^{-i\l\r} \ ,
\eeq
where $\r$ is the difference between  the quark and antiquark
impact parameters.

\beq{19}
\Psi^T_{\ga^*} (\r,z)=\mp i e  Q\sqrt{z\bar z}\left[  \de_{\la, -\la^\prime}
\{(1-2z)\mp \la \} (\e_{\pm}\frac {\r}{r})K_1(rQ\sqrt{z\bar z}) \right]
\eeq

\beq{20}
\Psi^L_{\ga^*} (\r,z)=-2e  Q z\bar z   \de_{\la, -\la^\prime}
K_0(rQ\sqrt{z\bar z})
\eeq
In these  eqs. (\ref{19},\ref{20}) $K_{0,1}(mr)$
are the Bessel (MacDonald) functions.

We can represent the impact factor
describing the $\ga^* \to \rho$ transition (2.10) in the form
(Fig.2)
\beq{21}
J_{\ga^* \to \rho}(\k,\q)=\int \frac{d^2\r_1dz_1}{16\pi^3}
 \Psi_{\ga^*} (\r_1,z)
\Phi^{dipole}
(\r_1,\r_2,\k,\q,z_1,z_2)
\Psi^*_\rho (\r_2,z) \frac{d^2\r_2dz_2}{16\pi^3} \,
\eeq
where $\Phi^{dipole}
(\r_1,\r_2,\k,\q,z_1,z_2)$ is the Fourier transform of (\ref{11})

\beqar{22}
&
\Phi^{dipole}(\r_1,\r,\k,\q,z_1,z_2)=16\pi^3
\frac{\as\de ^{ab}}{N}\de (\r-\r_1)\de (z_1-z_2)
f(\k,\q,\r,z_1) \ ,
& \nn \\
&
f(\r,\k,\q,z)=
e^{i\q\r z}(1-e^{-i\k\r})(1-e^{-i(\q-\k)\r}) \ .
&
\eeqar

The expression for the  dipole impact factor is proportional to
$\de (\r-\r_1)$ which reflects the mentioned above property
that the interaction does not change the impact parameters
of the pair. The main part of the dipole impact factor is the
factor $f(\r,\k,\q,z)$. This factor tends to zero  if
the momentum of one of the $t-$ channel gluon ($\k$ or $\q-\k$)
or if the transverse separation between the quarks $\r$ vanishes.

Let us discuss now the  wave functions of a vector meson.
They can be constructed using the analogy
with the photon wave functions.
The wave function of the longitudinally polarized  photon
can be rewritten in the following way
\beqar{23}
&
\Psi^L_{\ga^*}(\l,z)\sim z\bar z {\Phi (M) \over
z\bar z} \ ,
& \nn \\
&
\Phi(M)=\frac{\displaystyle 1}{\displaystyle Q^2+M^2} \ , 
M^2={l^2 \over z\bar z} .
&
\eeqar
Where $M$ is  the invariant mass of the $q\bar q$ pair.
We shall adopt  the
natural assumption that the meson wave functions depend
on the invariant mass of $q \bar q$ pair.
But in the $\rho$ meson case the corresponding
$\Phi (M)$ should fall off  faster at large
$M$ as compared to the photon wave function.
Therefore
\beq{24}
\Psi_\rho ^L (\l,z)=-\frac{3}{2}\de_{\la,-\la^\prime}
f_\rho z\bar z \frac{\Phi (l^2/(z\bar z))}{z\bar z} \ .
\eeq
The dimensionful coupling constant $f_\rho \sim 200 \mbox{ GeV} $ is related with
the $e^+e^-$ decay width of the $\rho $ meson,
\beq{25}
\Ga =\frac{2\pi \al^2 f_\rho^2}{3m_\rho} \ , \al =e^2/4\pi=1/137 \ .
\eeq
The normalization for function $\Phi(M)$ is
\beq{26}
\int \frac{d^2 l}{(16\pi^3)z\bar z}\Phi (\frac{l^2}{z\bar z})=1 \ .
\eeq

The wave function of the transversly polarized $\rho$
meson
\beq{27}
\Psi_\rho ^T (\l,z)=\pm \frac{3}{4}\de_{\la,-\la^\prime}
f_\rho \sqrt{z\bar z} \{(1-2z)\mp \la \} (\e_{\pm}\frac {\l}{l})
\frac{\Phi (l^2/(z\bar z))}{z\bar z}
\eeq
is constructed in analogy with the wave function of the
transverse photon. In the meson rest frame
the meson  splitting looks like as follows:
the quark and the antiquark fly out back to back, the polar angles of the quark
momentum relative to the direction of
the meson momentum in the boosted frame are $(\theta, \phi)$. The
amplitude of the transition of a spin 1 $\rho$--meson  in a state
with the helicity $\la$ into
the $q\bar q$ pair with the total helicity $\la^\prime$ is proportional
to the rotation matrix $d^1_{\la,\la^\prime}(\theta, \phi)$.
We are interested in  such meson splittings when the quark and antiquark
have the opposite helicities (zero total helicity) in the boosted frame, that corresponds
to the total helicity of the pair  $\pm 1$ in the rest frame.
Since the angle
$\theta$ in the rest frame and the variable $z$ in the boosted frame
 (the fraction of meson momentum
carrying by the quark) are related as
$$
z=\frac{1+\cos(\theta)}{2} \ ,
$$
it can be seen that the ratio of $\Psi_\rho ^L (\l,z)$ and
$\Psi_\rho ^T (\l,z)$ is equal to the ratio of the corresponding
rotation matrices. This observation justifies  eq. (\ref{27}) for
the wave functions of transverse $\rho$ as well as
the relative sign between the transverse and longitudinal wave functions.

The meson wave functions in the representation of the impact parameters
are given by the Fourier transform of eqs. (\ref{24},\ref{27}). Since
the $q\bar q$ fluctuation of the transverse polarized meson
has the projection $\pm 1$  of the angular moment onto the $z$
axis its wave function has a factor $(\e_\pm \r)$.

Now let us return  to  eq. (\ref{21}) and discuss the virtual
photon to meson impact
factor
\beq{28}
J_{\ga^* \to \rho}(\k,\q)=\frac{\as\de ^{ab}}{N}
 \int \frac{d^2\r dz}{16\pi^3}
 \Psi_{\ga^*} (\r,z)f(\r,\k,\q,z)
\Psi^*_\rho (\r,z)
\eeq
for the various helicity transitions.

Looking at the  eqs. (\ref{24},\ref{27},\ref{19},\ref{20})
for the meson and photon wave functions
it is easy to see that: \\
a)
the transitions obeying SCHC  are proportional to
the dipole factor  $f(\r,\k,\q,z)$ averaged over the polar angle
of the vector  $\r$
\beq{29}
\big< f(\r,\k,\q,z) \big> =[J_0(rqz)
-J_0(r|\k-\q z|)]+[z\leftrightarrow \bar z] \ ,
\eeq
b)
the single spin--flip  transitions are proportional to the projection of
this dipole factor onto $\frac{\displaystyle \r}{\displaystyle r}$
\beq{30}
\big< \frac{\r}{r} f(\r,\k,\q,z) \big> =
-i\left[
\frac{\q}{q} J_1(rqz)
+\frac{\k-\q z}{|\k-\q z|}J_1(r|\k-\q z|) -
(z\leftrightarrow \bar z)
\right] \ ,
\eeq
c)
the double spin--flip transitions
are related to the projections of this dipole factor onto ${{\bf r}_\mu
{\bf r}_\nu \over r^2 }$,  $\big< \frac{\r_\mu \r_\nu}{r^2}
f(\r,\k,\q,z) \big>$. They are proportional to
\beq{31}
\left[
2\frac{(k-\q z , \q)^2}{(\k-\q z)^2}J_2(r|\k-\q z|)
-q^2\left( J_2(r|\k-\q z|)
+ J_2(rq z)
\right)
\right] + \left[ z\leftrightarrow \bar z \right] \ .
\eeq

Using the general  formulae derived in this section we shall
calculate  in the next section the helicity amplitudes in the high
$Q^2$ limit
using the approximations to be described in the following.

\section{Calculating the helicity amplitudes}
\setcounter{equation}{0}

Assuming some functional form for the meson wave function
allows to calculate according to the eq. (\ref{28})
the $\ga^*\to \rho$ impact factor
without any approximation. Unfortunately the meson wave functions
are purely known. But, as it will be seen further, if we proceed
to calculate the amplitude in the leading order of $\frac{1}{Q}$
expansion only a limited information about these wave functions is needed.

The eqs. (\ref{19},\ref{20}) show that the
typical size of the virtual photon fluctuation
is $r \sim \frac{1}{Q\sqrt{z\bar z}}$. Outside of the end point
regions of z this size is much smaller than the meson transverse size.
Therefore to calculate the leading in  $ \frac{1}{Q} $
behaviour of the impact factor  we have to calculate
the  first term of the Taylor expansion for the meson wave function and
the expansion of the dipole factor $f(\r,\k,\q,z)$
in the region of small $\r$.

Making the Fourier transform of eqs. (\ref{24},\ref{27})
and then  expanding them at small $\r$ we find
\beq{32}
\Psi_\rho ^L (\r,z)\approx -\frac{16\pi^3}{2\pi}\frac{3}{2}
\de_{\la,-\la^\prime}
f_\rho z\bar z  \ ,
\eeq
\beq{33}
\Psi_\rho ^T (\r,z)\approx \mp i \frac{16\pi^3}{2\pi}
\frac{3}{8}\de_{\la,-\la^\prime}
f_\rho {z\bar z} \{(1-2z) \mp \la \} (\e_{\pm} {\r})\big< M \big> \ ,
\eeq
where
\beq{34}
\big< M \big>=\int \frac{d^2 l}{(16\pi^3)z\bar z}
\frac{l}{\sqrt{z\bar z}}\Phi (\frac{l^2}{z\bar z})
\eeq
is the mean invariant mass of the $q\bar q$ fluctuation. This mean invariant
mass is expected to be of order of the $\rho$ meson mass.
We would like  to note that the Taylor expansion for the transversly
polarized meson
wave function starts from the term $\sim r$,
whereas the longitudinal polarization wave function is constant at
small $r$.
The appearance of this suppression factor $r$ is related with the
nonzero projecton of the orbital momentum ($\pm 1$) of the quark pair in
the transverse case.

The azimuthal projections of the dipole factor $f(\r,\k,\q,z)$ for the
various
helicity transitions , see eqs. (\ref{29}--\ref{31}), can be simplified
in the region of small $\r$   by expanding  the Bessel functions. \\
a) for the SCHC transitions: \\
($J_0(x)\approx 1- \frac{x^2}{4}$)
\beq{35}
\big< f(\r,\k,\q,z) \big> =\frac{r^2}{2}(k^2-(\k\q)) + O(r^4) \ ;
\eeq
b) for the single spin--flip transitions: \\
($J_1(x)\approx \frac{x}{2}-\frac{x^3}{2^4}$) \\
\beq{36}
\big<\frac{\r}{r} f(\r,\k,\q,z) \big> =i\frac{r^3(z-\bar z)}{2^4}\left[
\k(q^2-2(\k\q))-\q(k^2-2(\k\q))
\right] + O(r^5) \ ;
\eeq
c) for the double spin--flip transitions:\\
($J_2(x)\approx \frac{x^2}{8}-\frac{x^4}{8\cdot 12}$) \\
\beqar{37}
& \frac{\displaystyle r^2}{\displaystyle 4q^2}\left[
2(\k\q)^2-(\k\q) q^2-k^2q^2 \right] & \nn \\
&
 -\frac{\displaystyle r^4}{\displaystyle 4\cdot 12q^2}
\left[ 2k^2(\k\q)^2-k^4q^2-2(\k\q)^3+ \right. & \nn \\
& \left.    3(\k\q)^2q^2(z^2+\bar z^2)-
2(\k\q) q^4(z^3+\bar z^3)  \right] + O(r^6) \ .&
\eeqar
We quote  in the last  case the expansion up to the
next-to-leading term,
the importance of which
 will be discussed in what
follows.

Let us discuss now  the proton impact factor and
the integration over the $t$-
channel gluon momenta in the eq. (\ref{9}).

The proton is a colorless state. Therefore its impact factor
vanishes
if the transverse momentum of any of the $t$- channel gluon tends to
zero. On the other hand if the transverse momenta of gluons are large,
much larger than  the inverse transverse size of the proton
(the value of the momentum transfer q is expected to be small,
therefore $\k\approx \k-\q$ in this region), both of the
gluons couple to the same parton inside the proton and as a function of
$\k$ the impact
factor is approximately a constant in this region.
According to eqs. (\ref{35},\ref{36}) the impact factors of the helicity
conserving
and single helicity flip $\ga^*\to \rho$ transitions are proportional
to the square of the $t$- channel gluon momentum $k^2$ at large $\k$:\\
a) for the SCHC transitions \\
\beq{38}
\big< f(\r,\k,\q,z) \big> \approx \frac{r^2k^2}{2} \ ;
\eeq
b) for the single spin--flip transitions \\
\beq{39}
\big<\frac{\r}{r} f(\r,\k,\q,z) \big> \approx
-i\q \frac{r^3k^2(z-\bar z)}{2^3} \ .
\eeq
Therefore the main, logarithmic, contribution to the helicity
non--flip and helicity single--flip amplitudes
originates from the broad region of large $\k$,
$k\leq \tilde Q=\frac{1}{Q\sqrt{z\bar z}}$. In the leading $\log \tilde Q^2$
approximation and at $t = 0$ these amplitude are proportional
to the gluon distribution
\beq{40}
x\cdot G(x,\tilde Q^2)=\frac{\de^{ab}}{2\pi}
\int\limits^{\tilde Q^2}_{} J_p^{ab} \frac{d k^2}{k^2} \ .
\eeq

Let us calculate these amplitudes in the leading log approximation.

\subsection{SCHC amplitudes}

Let us start with the dominant at large $Q^2$
longitudinal amplitude.
Using (\ref{40}) and
inserting eqs. (\ref{20},\ref{32},\ref{38}) into eq. (\ref{28})
and performing the sum over
the quark helicities we obtain the following expession
\beq{41}
M_{(0,0)}=is_{\ga^* p} \int d^2r dz r^2
\frac{3\pi e\as f_\rho}{\sqrt{2}N}Q(z\bar z)^2K_0 (rQ\sqrt{z\bar z})
x G(x, Q^2 z\bar z) \ .
\eeq
We took into account that the mean electric charge of the quarks
inside the $\rho$ meson
($|\rho\big>=\frac{1}{\sqrt{2}N}(|u\bar u\big>-|d\bar d\big>)$) is $e/\sqrt{2}$.
Performing the integral over $ \r$, using $\int\limits^\infty_0 K_0(r)
r^3 dr=4$, we obtain
the result
\beq{42}
M_{(0,0)}=is_{\ga^* p} \int dz
\frac{8\cdot3\pi^2 e\as f_\rho}{\sqrt{2}NQ^3}
x G(x, Q^2 z\bar z) \ .
\eeq
If we neglect the $z$ dependence of the argument of the gluon density
and take the integral over $z$ in eq. (\ref{42})
we will reproduce the known result for the longitudinal amplitude
\cite{r,br},
( for the  asymptotical form of the meson distribution amplitude).

The integral over $z$ is convergent in eq. (\ref{42}). Therefore the end
point regions of $z$ do not bring the essential contributions,
and typical transverse distances of the process are small,
$\sim \big< \frac{1}{Q\sqrt{z\bar z}} \big>$.
This justifies the  application of the perturbative QCD  in this case.

The situation is more complicated for the case of the
transverse amplitude. Inserting eqs. (\ref{19},\ref{33},\ref{38})
into (\ref{28}) we obtain
\beqar{43}
& M_{(+1,+1)}=is_{\ga^* p} \int d^2r dz r^3
\frac{\displaystyle 3\pi e\as f_\rho}{\displaystyle 8N\sqrt{2}} & \nn \\
& \big<M\big>Q(z\bar z)^{3/2}(z^2+\bar z^2)
K_1 (rQ\sqrt{z\bar z})
x G(x, Q^2 z\bar z) \ . &
\eeqar
The integration over $\r$ gives, using $\int\limits^\infty_0 K_1(r) r^4 dr=16$,
\beq{44}
M_{(+1,+1)}=is_{\ga^* p} \int dz
\frac{4\cdot 3\pi^2 e\as f_\rho \big<M\big>
}{\sqrt{2}NQ^4(z\bar z)}(z^2+\bar z^2) x G(x, Q^2 z\bar z) \ .
\eeq
If we would neglect the scale dependence of the gluon density $G(x, \bar
Q^2=Q^2 z\bar z)$ the integration over $z$
would be  logarithmically divergent at $z\to 0,1$ in the above integral.
On the other hand
in the small $x$ region the gluon density increases rapidly with
$\bar Q^2$. This increase can be simulated in a first approximation as
\beq{444}
G(x, Q^2 z\bar z)=G(x, Q_0^2) \left[(Q^2z\bar z)/Q_0^2 \right]^\ga \ ,
\eeq
with a constant anomalous dimension $\gamma $
of the gluon density. Then the integral over $z$ in the eq. (\ref{44})
is convergent and we confirm  the result of \cite{mrt}
\beq{45}
\al =\frac{M_{(+1,+1)}}{M_{(0,0)}}=\frac{ \big<M\big>}{Q}\frac{1+\ga}{\ga} \ .
\eeq

\subsection{The helicity single--flip amplitudes}

The calculations  of the two independent
helicity single--flip amplitudes are  quite
similar to the ones for the SHCH amplitudes.

Using eqs. (\ref{19},\ref{32},\ref{39})
we obtain the following expression
for the single spin--flip transition of the transversly
polarized initial photon

\beqar{46}
& M_{(+1,0)}=is_{\ga^* p} \int d^2r dz r^3
\frac{\displaystyle 3\pi e\as f_\rho}{\displaystyle 16N} & \nn \\
& \sqrt{|t|}Q(z\bar z)^{3/2}(z-\bar z)^2
K_1 (rQ\sqrt{z\bar z})
x G(x, Q^2 z\bar z) \ .&
\eeqar
The integration over $\r$ gives
\beq{47}
M_{(+1,0)}=is_{\ga^* p} \int dz
\frac{2\cdot 3\pi^2 e\as f_\rho \sqrt{|t|}
}{NQ^4(z\bar z)}(z-\bar z)^2 x G(x, Q^2 z\bar z) \ .
\eeq

According to eqs. (\ref{20},\ref{33},\ref{39})
the single spin--flip transition of the longitudinaly
polarized initial photon is

\beqar{48}
& M_{(0,+1)}=-is_{\ga^* p} \int d^2r dz r^4
\frac{\displaystyle 3\pi e\as f_\rho}{\displaystyle 32N} & \nn \\
& \sqrt{|t|} \big<M\big>Q(z\bar z)^2(z-\bar z)^2
K_0 (rQ\sqrt{z\bar z})
x G(x, Q^2 z\bar z) \ .&
\eeqar
After the integration over $\r$, using $\int\limits^\infty_0
 K_0(r) r^5 dr=64$, we have
\beq{49}
M_{(0,+1)}=-is_{\ga^* p} \int dz
\frac{4\cdot 3\pi^2 e\as f_\rho \sqrt{|t|}\big<M\big>
}{NQ^5(z\bar z)}(z-\bar z)^2 x G(x, Q^2 z\bar z) \ .
\eeq

We calculate the integrals over $z$ in (\ref{42},\ref{47},\ref{49})
using the
assumption of the constant gluon anomalous dimension (\ref{444}).
The following results for the ratios of the
single--flip amplitudes to the longitudinal non--flip
amplitude can be derived

\beq{50}
\be =\frac{M_{(+1,0)}}{M_{(0,0)}}=\frac{ \sqrt{|t|}}{Q}\frac{1}{\sqrt{2}\ga}
\ ,
\eeq

\beq{51}
\de =\frac{M_{(0,+1)}}{M_{(0,0)}}=-\frac{ \big<M\big> \sqrt{|t|}}{Q^2}
\frac{\sqrt{2}}{\ga} \ .
\eeq

\subsection{The helicity double--flip amplitude}

The helicity double spin--flip amplitude $M_{(+1,-1)}$ can be written
as the sum of the two parts
\beq{52}
M_{(+1,-1)}=M^0_{(+1,-1)}+M^1_{(+1,-1)} \ ,
\eeq
where $M^0_{(+1,-1)}$ corresponds to the first ($\sim r^2$)
term of eq. (\ref{37}), and  $M^1_{(+1,-1)}$ corresponds to the second
($\sim r^4$) term of  eq. (\ref{37}). Keep in mind that the expansion in
$r$ is equivalent to the expansion of the amplitude in
$\frac{1}{Q\sqrt{z\bar z}}$.

The integrations over the transverse momenta of the $t-$  channel
gluons, see eq. (\ref{9}), and over the $q\bar q$ longitudinal
momentum fraction z  are diffrerent for $M^0_{(+1,-1)}$ and
$M^1_{(+1,-1)}$.

Let us discuss  $M^0_{(+1,-1)}$ first.
Using (\ref{19},\ref{33},\ref{37}) we obtain
\beq{53}
M^0_{(+1,-1)}=-is_{\ga^* p} \int \frac{d^2r dz r^3}{2\pi}
\frac{3 e f_\rho}{8N\sqrt{2}}
 \big<M\big>Q(z\bar z)^{5/2}
K_1 (rQ\sqrt{z\bar z})
I_0 \ ,
\eeq
where the factor $I_0$ represents the integration over the transverse
momenta of the $t-$ channel gluons
\beq{54}
I_0=\frac{\as \de^{ab}}{q^2}\int \frac{d^2 k}{\k^2 (\k -\q)^2}
\left[2(\k\q)^2-k^2q^2-(\k\q)q^2\right]
J_p^{ab}(\k,\q) \ .
\eeq
Performing the integration over $\r $  in (\ref{53})
we have
\beq{55}
M^0_{(+1,-1)}=is_{\ga^* p} \int dz
\frac{ 3\cdot 2 e f_\rho  \big<M\big> }{\sqrt{2}N Q^4}
I_0 \ .
\eeq
The above equation should be compared with the corresponding eq. (\ref{44})
for the helicity non--flip transverse amplitude $M_{(+1,+1)}$.
In contrast to (\ref{44}), the integration over $z$ in (\ref{55})
is not singular in the end point regions. This difference is related
with the helicity flip. Looking at the
expressions (\ref{19}) and (\ref{33}) for the
photon and vector meson wave functions it is seen that
in the flip case the sum over the helicities of the intermediate
$q\bar q$ pair is proportional
to the additional  factor  $z\bar z$.
For the non--flip case the sum
over $q\bar q$ helicities  gives the factor $\sim (z^2 + \bar z^2) $,
which does not vanish at  $z =0,1$.

Now let us discuss  the integration over the $t-$ channel
gluon momenta (\ref{54}).
This integral is convergent on the upper limit.
Therefore  $M^0_{(+1,-1)}$ can not be expressed through
the gluon density using  eq. (\ref{40}). Moreover,
since the main part of the integral (\ref{54}) originates from the region
of small transverse momenta of the $t-$ channel gluons,
$k\sim q$, we are dealing here with soft physics.

We shall assume that this soft $t-$ channel exchange can be
described  in the frame of the  two--gluon
exchange model, see eq. (\ref{9}), with some simple functional form for the
nonperturbative impact factor of the proton $J_p$. The model of
such type was used successfully in \cite{gs} to describe the total
$pp$ cross sections. Following \cite{gs}, the proton impact factor is
\beq{56}
J_p(\k,\q)=\bar \as \de^{ab} \left[
\frac{A^2}{A^2+q^2/4}-\frac{A^2}{A^2+(\k-\q/2)^2}
 \right] \ .
\eeq
The first term in (\ref{56}) describes the contribution of the
diagrams where both  $t-$ channel gluons are coupled to the
same quark inside the proton. This contribution is similar
to the electromagnetic form factor, therefore it is natural
to adopt that
\beq{57}
A=\frac{m_{\rho}}{2} \ .
\eeq
The strength of the nonperturbative coupling $\bar \as=\frac{\bar g^2}{4\pi}$
is a free parameter. To choose its value, let us calculate
the value of the total $pp$ ($p\bar p$) cross section which is related through
the optical theorem to the forward amplitude
\beq{58}
\si_{tot}^{pp}=\frac{Im (M^{pp}(t=0))}{s_{pp}} \ .
\eeq
Using eqs. (\ref{9}) and (\ref{54}) it can be shown that
\beq{59}
Im (M^{pp}(t=0))=\frac{8\pi \bar \as^2 s_{pp}}{A^2} \ .
\eeq
Describing the $pp$ total cross section with this simple model
leads to an parameter
$\bar \as$
increasing with energy. We understand that one can do better
by replacing the two gluon exchange by a pomeron exchange.
For our aim of a simple estimate we shall adopt the relations
(\ref{58},\ref{59}).

Performing the integral (\ref{54}) (using the expression (\ref{56})
for the proton impact factor) we find
\beq{60}
I_0=8\pi\bar \as^2 \frac{A^2}{A^2+q^2/4}
\left[
(1+\frac{4A^2}{q^2})\log{(1+\frac{q^2}{4A^2})}-1
\right] \ .
\eeq
At small $q^2$, $q^2<<  A^2$,
\beq{61}
I_0=\pi\frac{q^2}{A^2}=\pi\frac{4|t|}{m_{\rho}^2} \ .
\eeq
Our final result for $M^0_{(+1,-1)}$ is
\beq{62}
M^0_{(+1,-1)}=-is_{\ga^* p}
\frac{ 3\cdot 8 \pi e \bar \as^2 f_\rho |t| \big<M\big>  }
{\sqrt{2}N Q^4 m_{\rho}^2}
\ .
\eeq
Perfoming the
integral over $z$ in (\ref{42}) we find
\beq{622}
\eta^0=\frac{M^0_{(+1,-1)}}{M_{(0,0)}}=
-\frac{\bar \as^2 |t|\big<M\big>}{\pi \as Q m^2_
{\rho^0}}\frac{1}{\left[ 4^\ga\frac{\Ga^2 (\ga+1)}{\Ga (2\ga +2) }
xG(x,Q^2/4) \right]} \ .
\eeq

Now let us discuss $M^1_{(+1,-1)}$.
Using (\ref{19},\ref{33}) and the second term of (\ref{37})
we can write
\beq{63}
M^1_{(+1,-1)}=is_{\ga^* p} \int \frac{d^2r dz r^5}{2\pi}
\frac{3 e f_\rho}{8\cdot 12N\sqrt{2}}
 \big<M\big>Q(z\bar z)^{5/2}
K_1 (rQ\sqrt{z\bar z})
I_1 \ .
\eeq
In this case the integration over the transverse
momenta of the $t-$ channel gluons is different
as compared to $M^0_{(+1,-1)}$ case
\beqar{64}
& I_1=\frac{\as \de^{ab}}{q^2}\int \frac{d^2 k}{\k^2 (\k -\q)^2}
\left[
2k^2(\k\q)^2-k^4q^2-2(\k\q)^3+  \right.& \nn \\
& \left. 3(\k\q)^2q^2(z^2+\bar z^2)-
2(\k\q) q^4(z^3+\bar z^3)
\right]
J_p^{ab}(\k,\q) \ . &
\eeqar
Extracting the main, logarithmic, part from the above integral
we find
\beq{65}
I_1=\frac{\as \de^{ab}}{q^2}
\frac{q^4\left[3(z^2+\bar z^2)-1\right]}{2}
\int \frac{d^2 k}{k^2 }
J_p^{ab}(\k,\q) \ .
\eeq
Therefore we can relate
$M^1_{(+1,-1)}$, using the relation (\ref{40}),  to the
gluon density.
Performing the integral over $\r$,
using $\int\limits^\infty_0
 K_1(r) r^6 dr=384$, we have
\beq{66}
M^1_{(+1,-1)}=is_{\ga^* p} \int dz
\frac{4\cdot 3\pi^2 e\as f_\rho |t|\big<M\big>
}{\sqrt{2}NQ^6}\frac{\left[3(z^2+\bar z^2)-1\right]}
{(z\bar z)} x G(x, Q^2 z\bar z) \ .
\eeq
Note that the $z$ integration is different for $M^1_{(+1,-1)}$ and
$M^0_{(+1,-1)}$. The additional factor $(z\bar z)Q^2$
appears in the denominator of (\ref{66}) by virtue of
two additional powers of $r$ in the numerator of eq. (\ref{63}).
As a result the $z$ integration for $M^1_{(+1,-1)}$ becomes
similar to ones for $M^1_{(+1,+1)}$, $M^1_{(0,+1)}$ and $M^1_{(+1,0)}$.
Calculating the integrals over $z$ ( using (\ref{444}) in eqs. (\ref{42})
and (\ref{66})) we find
\beq{67}
\eta^1=\frac{M^1_{(+1,-1)}}{M_{(0,0)}}=
\frac{ |t|\big<M\big>}{Q^3}\frac{2(\ga + 2)}{\ga} \ .
\eeq

\beq{68}
\eta = \frac{M_{(+1,-1)}}{M_{(0,0)}}=\eta^0+\eta^1 \ .
\eeq

It should be noted that although $M^1_{(+1,-1)}$
is suppressed as compared to $M^0_{(+1,-1)}$
by the factor $\sim \frac{m^2_\rho}{Q^2}$,
the perturbative part $M^1_{(+1,-1)}$ can be dominant 
at high energies since it has a steeper energy dependence 
($\sim xG(x,\bar Q^2)$) as compared to the soft part 
$M^0_{(+1,-1)}$. $M^1_{(+1,-1)}$ contains also an enhancment
factor $1/\ga$ which originates from the more singular 
$z$ integration. We will show below that 
$|M^1_{(+1,-1)}|>M^0_{(+1,-1)}$ at typical for HERA
kinematical conditions.

\subsection{Additional remarks on the amplitudes}

We have calculated above the  helicity amplitudes for the
diffractive vector meson electroproduction at $s_{\ga^*}>>Q^2>>\La_{QCD}^2$
(large $Q^2$, small $x$).
Let us discuss
now the assumptions used and the physical issues
related with the helicity amplitudes.

We assume that perturbative QCD can be applied to
describe the process (\ref{1}) at large $Q^2$.
Although the QCD factorization theorem has been proven
only for the leading scalar amplitude $M_{(0,0)}$,
we extend here the perturbative
aproach, following \cite{mrt} (where transverse helicity non--flip
amplitude $M_{(+1,+1)}$ was considered),  to
describe helicity--flip amplitudes.

The main features of the perturbative QCD aproach are the 
following. The virtual photon splits into the massless $q\bar q$ pair 
having the total helicity 0. As a result,  the helicity of the photon 
coincides with the $z$ projection of the quark 
angular momentum.  The helicity states of the quarks do not 
change during the interaction with the proton. Therefore the helicity 
of the meson is equal to the projection of the angular momentum of 
the outgoing $q\bar q$ pair onto the direction of the meson momentum.
The helicity flip  
comes from the change of the projection of the $q\bar q$ angular
momentum during the interaction.  
This change 
originates in our aproach from the non--forward
kinematics ($t\neq 0$). Therefore there is no suppression
of the  
helicity--flip amplitudes with energy, they are driven  
by the leading gluon (pomeron) exchange. $M_{(+1,0)}$, $M_{(0,+1)}$
and $M^1_{(+1,-1)}$ have the energy dependences which are
similar to the energy dependences of the helicity
non--flip amplitudes. They are proportional (at small $t$) 
to $xG(x,\bar Q^2)$.

According to our calculations the helicity  single--flip amplitudes are 
proportional to $\sqrt{|t|}$, the double spin--flip one is 
$\sim |t|$. These factors come from the expansion of the 
$\ga^*\to \rho$ transition impact factors describing these amplitudes. 
Since the typical transverse separation between the quarks is
small, $\sim \frac{1}{Q\sqrt{\ga}}$, this expansion is expected 
to be valid up to the rather large values of $|t|$, $|t|\leq Q^2\ga$.
On the other hand, equation (\ref{40}) relating the proton impact factor 
with the gluon density and  its implications 
eqs. (\ref{42},\ref{44},\ref{47},\ref{49},\ref{66}) 
are  valid only at very small (vanishing)
momentum transfer. The coupling of the two gluon system with the 
proton decreases with the growth of the momentum transfer.
We are not able to describe  the $t$ behaviour
of this coupling
from the first principles. But we see that at small $t$
the structure of the integration over the transverse momenta 
of the $t-$ channel gluons is similar for  both 
the helicity non--flip 
and the helicity--flip amplitudes (with the exception of the nonperturbative 
part  of the double--flip amplitude  $M^0_{(+1,-1)}$). Therefore it is natural 
to expect an universal $t$ behaviour for all helicity amplitudes coming 
from the coupling of the $t-$ channel gluons with the proton.
And we believe  that this universal $t$ dependence is canceled in the
ratios (\ref{45},\ref{50},\ref{51},\ref{67}) and, therefore, these results
are valid 
in the broad $t$ region extending up to $|t|\leq Q^2\ga$. 
This assumption is based on the observation that in all cases we deal with 
the scattering of a $q\bar q$ pair the transverse size of which
is much smaller than 
the size of the proton. 

We will 
parametrize this universal $t$ 
behaviour as $\sim e^{-b|t|/2}$ for the amplitudes 
($\sim e^{-b|t|}$ for the cross sections) 
with the slope $b$ which is of the order 
of the square of the  proton size.
Due to the large value of this slope, $b=5\dots 6 \mbox{ GeV}^{-2}$
according to the HERA data \cite{c}, the helicity flip amplitudes are peaked
as well as the helicity non--flip ones at  small $t$ ($|t|\leq t_0\sim 1/b$)
in spite of the fact that they contain the factors $\sqrt{|t|}$ 
for single--flip
and $|t|$ for double--flip.

According to eqs. (\ref{45},\ref{50},\ref{51},\ref{68}) 
for the typical $|t|$ values , $|t|\approx 1/b$, 
assuming that $\big< M\big>\sim m_\rho$, we have
\beq{69}
1>\al>\be>|\de|>|\eta| \ .
\eeq
Therefore the largest among the amplitudes 
violating SCHC, the 
helicity single--flip amplitude $M_{(+1,0)}$, is smaller than
the transverse non-flip amplitude $M_{(+1,+1)}$.   

Let us  estimate the  ratios in eq. (\ref{69})
for the kinematical conditions relevant  for the HERA experiments.  
We choose $Q^2=10 \mbox{ GeV}^2$, $x=10^{-3}$ which corresponds to 
$W=\sqrt{s_{\ga^*p}}=100 \mbox{ GeV}$. We will give the estimates 
for the simplest situation when the momentum transfer is not
restricted during the helicity analysis, i.e. the experimental sample
is not divided into the $t$ bins. In this case we can substitute the 
factors $\sqrt{|t|}$ and $|t|$ by their mean values
$$
|t|\to \big< |t|\big>=\frac{\displaystyle 
\int dt |t| e^{-b|t|}}{\displaystyle \int dt  e^{-b|t|}}
=\frac{1}{b} \ ,
$$  
$$
\sqrt{|t|}\to \big< \sqrt{|t|}\big>=
\frac{\displaystyle \int dt \sqrt{|t|} e^{-b|t|}}{\displaystyle 
\int dt  e^{-b|t|}}=
\frac{\sqrt{\pi}}{2\sqrt{b}}
 \ .
$$
We shall use $b=6  \mbox{ GeV}^{-2}$, $\big< M\big>= m_\rho$ 
in our estimates.
For the effective gluon anomalous dimension  at these values 
of $Q^2$ and $x$ we use two values 
$\ga =0.7$ and $\ga =0.5$. These values of $\ga$ are close to the ones 
presented in 
Fig. 3 of \cite{mrt}. 

The results of our estimates for $\ga =0.7$ ($\ga=0.5$) are the following:
\beq{70}
\al= 0.59  (0.73)\ ,
\eeq
\beq{71}
\be= 0.12 (0.16)\ ,
\eeq
\beq{72}
\de= 0.056 (0.079)\ ,
\eeq
\beq{73}
\eta^1=0.031 (0.041) \ ,
\eeq
\beq{74}
\eta^0=-0.016 (-0.015) \ .
\eeq
Estimating $\eta^0$ in eq. (\ref{622}) we use 
$\as=0.3$, $\bar \as=0.87$. This value of $\bar \as=0.87$
used in eq. (\ref{59}) reproduces the correct value of the 
total $pp$ cross section at $\sqrt{s_{pp}}=100\mbox{ GeV}$. 
Combining eqs. (\ref{73},\ref{74}) we obtain a very small 
number for the ratio of $M_{(+1,-1)}$ to  $M_{(0,0)}$
\beq{75}
\eta=0.015 (0.026) \ .
\eeq
$M_{(+1,0)}$ is approximately $7\dots 8 $ times smaller than  
$M_{(0,0)}$. The other helicity--flip amplitudes are considerably
more suppressed.

We have calculated above the dominant at high energy imaginary 
parts of the helicity amplitudes. 
We will give here only an argument  
why the real parts can  be not too important in the 
polarization phenomena. 
The real parts of the amplitudes 
are related to the imaginary ones through 
the dispersion relations. Since the imaginary parts
of all helicity amplitudes have 
according to our consideration  similar energy behaviour
we expect that the helicity amplitudes have  phases which are 
close to each other. Therefore  
observable effects related to the 
differences of these
phases  will be additionally 
suppressed.

\section{Vector meson decay angular  distribution
at HERA kinematics}
\setcounter{equation}{0}

The polarization of the $\rho$ meson is experimentally accessible 
through the measurement of  the angular distributions of the decay products. 
For the relations between the helicity 
amplitudes and the angular distributions 
we shall use the results and standard 
conventions of \cite{sw}, see also \cite{c}. 
The definition of the three 
independent angles involves three planes:
1) the electron scattering plane, 2) the vector meson production plane
(which contains the photon and meson  momentum vectors),
3) the meson decay plane. The orientation of the meson decay 
plane is described by the polar and the azimuthal angles $(\theta )$ and 
$(\phi )$.
The third angle $(\Phi )$ is the angle between the electron scattereing
and the meson production planes. 

The polarization parameter of the virtual photon density matrix, 
$$
\epsilon = \frac{1-y}{1-y + \frac{\displaystyle y^2}{\displaystyle 2}} \ ,
$$
is  close to 1 at HERA kinematics. For $W=100\mbox{ GeV}$ 
($y=\frac{\displaystyle s_{\ga^*p}}{\displaystyle s_{ep}}=1/9$) 
its value is $\epsilon =0.993$.

The decay distribution $W(\cos{\theta},\phi,\Phi)$ contains
the parameters, the  matrix elements  $r^\al_{ik}$, 
which are  known bilinear 
combinations of the helicity amplitudes \cite{sw}. These matrix elements
can be determined experimentally by the  analysing moments
of the observed decay 
angular distribution.

According to our estimate  
the helicity--flip amplitudes are substantially
smaller than the helicity non--flip ones. Nevertheless, 
as we shall show, the largest among them $M_{(+1,0)}$ leads to a sizable 
effect.

The ratio of the longitudinal to the transverse cross section is 
expressed through the ratios of the helicity amplitudes as follows
\beq{76}                                                         
R=\si_L/\si_T=N_L/N_T \ ,               
\eeq
where
\beq{77}
N_T=\al^2+\be^2+\eta^2 \ ; N_L=1+2\de^2 \ . 
\eeq 
The matrix elements entering  the 
decay angular distribution  
have the following expressions  in terms of the ratios of the helicity 
amplitudes ( 
assuming that the  helicity amplitudes are purely imaginary):     
\beqar{78}
&
r^{04}_{00}=B(\epsilon+\be^2) \ , \ 
Re (r^{04}_{10})=B(2\epsilon\de+\be\al-\be\eta)/2 \ , \
r^{04}_{1-1}=B(\al\eta-\epsilon\de^2) \ ; 
& \nn \\
&
r^{1}_{11}=B\al\eta \ , \
Re (r^{1}_{10} )=B\be(\eta-\al)/2 \ , \
r^{1}_{00}=-B\be^2 \ ,
r^{1}_{1-1}=B(\al^2+\eta^2)/2 \ ;
& \nn \\
&
Im (r^{2}_{10} )=B\be(\al\ +\eta)/2 \ , \
Im (r^{2}_{1-1} )=B(\eta^2- \al^2)/2 \ ;
& \nn \\
& 
r^{5}_{11}=\frac{\displaystyle B}{\displaystyle \sqrt{2}}
\de (\al-\eta) \ , \
Re (r^{^5}_{10})=\frac{\displaystyle B}{\displaystyle \sqrt{2}}
(2\be\de+\al-\eta)/2 \ , 
& \nn \\
&
r^{5}_{00}=\frac{\displaystyle B}{\displaystyle \sqrt{2}}
2\be \ , \
r^{^5}_{1-1}=\frac{\displaystyle B}{\displaystyle \sqrt{2}}
\de (\eta-\al) \ ;
& \nn \\
&
Im (r^{6}_{10})=-\frac{\displaystyle B}{\displaystyle \sqrt{2}}
(\al+\eta)/2 \ , \
Im (r^{6}_{1-1})=\frac{\displaystyle B}{\displaystyle \sqrt{2}}
\de (\al+\eta)/2 \ .
& 
\eeqar
We introduce for short the notation 
$$
B=1/(N_T+\epsilon N_L) \ .
$$

Substituting our estimates for the ratios of the helicity 
amplitudes derived in the previous section we have  
\beqar{79}
&
r^{04}_{00}= 0.74(0.65)\ , \ 
Re (r^{04}_{10})=-0.015(-0.014) \ , 
& \nn \\
&
r^{04}_{1-1}=0.0042(0.0082) \ ; 
& \nn \\
&
r^{1}_{11}=0.0065(0.012) \ , \
Re (r^{1}_{10} )=-0.025(-0.036) \ , 
& \nn \\
&
r^{1}_{00}=-0.011(-0.016) \ ,
r^{1}_{1-1}=0.13(0.17) \ ;
& \nn \\
&
Im (r^{2}_{10} )=0.027(0.039) \ , \
Im (r^{2}_{1-1} )=-0.13(-0.17) \ ;
& \nn \\
& 
r^{5}_{11}=-0.017(-0.025) \ , \
Re (r^{^5}_{10})=0.15(0.15) \ , 
& \nn \\
&
r^{5}_{00}=0.12(0.14)\ , \
r^{^5}_{1-1}=0.017(0.025) \ ;
& \nn \\
&
Im (r^{6}_{10})=-0.16(-0.17) \ , \
Im (r^{6}_{1-1})=-0.0088(-0.013) \ .
& 
\eeqar

The matrix elements which have nonzero values in the case of SCHC (
if $\be,\de,\eta =0$) are $r^{04}_{00}, r^{1}_{1-1},
Im (r^{2}_{1-1}), Re (r^{5}_{10}), Im (r^{6}_{10})$.
The relations which would hold between them in the 
the case of SHCH are only      slightly
violated since  $\de,\eta$ are very small: 
\beq{80}
\frac{1}{2}(1-r^{04}_{00})-r^{1}_{1-1}=B\epsilon\de^2\approx 0.002(0.004) \ , 
\eeq
\beq{81}
r^{1}_{1-1}+Im (r^{2}_{1-1})=B\eta^2\approx 2(4)\cdot 10^{-4} \ ,
\eeq
\beq{82}  
Re (r^{5}_{10})+Im (r^{6}_{10})=
\frac{\displaystyle B}{\displaystyle \sqrt{2}}(\be\de-\eta) 
\approx -0.011(-0.017)\ .                   
\eeq
Therefore these matrix elements should  
be measured with high precision to see the violation of 
the SCHC. Note also that the 
value of $R$ calculated using the SCHC relation 
$R=\frac{\displaystyle 1}{\displaystyle \epsilon}
\frac{\displaystyle r^{04}_{00}}{\displaystyle 1-r^{04}_{00}}$
exceeds the one calculated using eq. (\ref{76}) by $3\dots 4 \%$. 

It is natural that the effects of the violation of 
SCHC manifest themselves more 
transparent in other matrix elements which would be zero in the case 
of SCHC. The largest among them is $r^{5}_{00}\approx 0.12(0.14)$.  
This matrix element  has a very 
clear meaning. It is related to the interference between the
two amplitudes describing the two possibilities
to produce  the longitudinally polarized 
vector meson. These are    
the dominant
at large $Q^2$ amplitude $M_{(0,0)}$ and the helicity single--flip 
amplitude $M_{(+1,0)}$. 

Since this interference exists on the level 
of the  production of the vector meson and does not depend on the 
kinematical variables describing the meson decay ($\theta , \phi$),
the correspondig 
effect survives without the loose of 
the analysing power after the integration of the angular distribution
over the angles 
($\theta , \phi$). The resulting 
distribution over the relative angle between the electron 
scattering and the   meson production planes is 
\beq{83}
W(\Phi)= \left[1+\sqrt{2\epsilon (1+\epsilon)}\cos{\Phi}
(r^5_{00}+2r^5_{11})\right] \ .
\eeq 
We skipped in the above equation the term $\sim \cos{2\Phi}$
which is proportional to the small matrix elements $r^1_{00}, r^1_{11}$.
Substituting  our estimates for $r^5_{00}, r^5_{11}$ we obtain a
substantial deviation of the $\Phi$ distribution from the flat one
\beq{84}
W(\Phi)= \left[1+0.18(0.19)\cos{\Phi}\right] \ .
\eeq

The other matrix element that could be potentially large is 
$Re (r^{04}_{10})$. Since it contains the term 
$\sim 2\epsilon \de$
which is linear in $\de$, 
 this matrix element 
has a large  sensitivity to the second single--flip amplitude 
$M_{(0,+1)}$. But its value turns out to be small due to a large 
cancelation between the terms $ 2\epsilon \de$ and $\be\al$.
This cancelation is related with the opposite signs 
of $M_{(+1,0)}$ and $M_{(0,+1)}$. Note that in the case of positive 
sign of $\de$ this matrix element would be estimated as 
$Re (r^{04}_{10})=0.066(0.086)$.

The matrix elements (\ref{78})
parametrize the angular distribution for the  unpolarized
initial positron.  A few additional matrix elements
can be measured if the initial positron is longitudinally polarized.
But these matrix elements do not exhibit an advantage in  sensitivity to 
the violation of SCHC as compared to the unpolarized ones. 

\section{Conclusions}
\setcounter{equation}{0}

We have considered in perturbative QCD the polarization effects 
in  diffractive $\rho$ meson electroproduction.
We assume that perturbative QCD with the account of the important
effect of the gluon scale behaviour 
is applicable to all
helicity amplitudes. 
We estimate the helicity  amplitudes in the approximation of the 
constant effective anomalous dimension of the gluon.
Our results are summarized in the eqs. (\ref{45},\ref{50},\ref{51},
\ref{67},\ref{68}).

The equations derived here can be applied also to other light 
vector mesons $\omega, \phi$ with corresponding changes in the coupling 
constants and parameters $\big < M\big>$ describing the meson wave functions.

The perturbative QCD leads to a very definite qualitative picture
for the violation of SCHC at high $Q^2$. 
The largest among the helicity--flip amplitudes 
is $M_{(+1,0)}$. 
At typical values of 
$t$ this amplitude is smaller than the transverse helicity 
non--flip amplitude  $M_{(+1,+1)}$,
$M_{(+1,0)}\sim
\frac{\displaystyle  \sqrt{|t|}}{\displaystyle m_\rho}  M_{(+1,+1)}$. 
The other independent single--flip 
amplitude $M_{(0,+1)}$ is suppressed compared to 
$M_{(+1,0)}$ by the factor 
$\frac{\displaystyle 2m_\rho}{\displaystyle Q}$.
The double--flip amplitude $M_{(+1,-1)}$ consists of the  two parts.
The perturbative contribution $M^1_{(+1,-1)}$ is larger 
at small $x$ than the nonperturbative one. But it is suppressed 
compared to $M_{(0,+1)}$ by the additional factor $\sim 
\frac{\displaystyle \sqrt{|t|}}{\displaystyle Q}$.
Therefore at very high $Q^2$ we have 
$M_{(+1,0)}>>|M_{(0,+1)}|>>M_{(+1,-1)}$.
For the kinematical region typical for the HERA experiments 
we 
find that $|M_{(0,+1)}|$ is about two times smaller than  $M_{(+1,0)}$,
and $M_{(+1,-1)}$ is about 10 times smaller than   $M_{(+1,0)}$.

This hierarchy between the helicity amplitudes leads to the peculiar 
predictions for the parameters of the 
meson decay angular distribution.
We predict that the only one parameter 
among that vanishing in the case of SCHC, $r^5_{00}$,
deviates substantially
from zero. The relations between the parameters, which are 
nonzero in the case of  SCHC, are only slightly violated.
It would be very interesting to confront these predictions
with the data.
  
Let us note that the parameter $r^5_{00}\sim M_{(+1,0)}$
is sensitive to the meson wave function. Since 
single spin--flip transitions are proportional 
to the factor vanishing if $z=\bar z$, see (\ref{30},\ref{39}),
the helicity amplitude  $M_{(+1,0)}$ whould be zero if
the wave function of the meson is a non-relativistic one
$\sim \de (z-1/2)$.  It is a consequence of  perturbative QCD 
that the fluctuation of the light meson into the pair of  current
quarks is described by the broad wave function. On the other hand
in the nonperturbative models a meson is often 
considered as a weakly bounded system of 
constituent quarks having a mass $m_q\sim m_M/2$ and, 
therefore, described by a function close to $\de (z-1/2)$.
Therefore the large value of  $r^5_{00}$ is a 
characteristic 
prediction of perturbative QCD.   
      
Note that 
our work is only a first step in the investigation of 
the helicity--flip amplitudes.  
More detailed numerical calculations  can be done using the  equations 
derived in this paper. It is possible to do this without
the approximation of the 
constant gluon anomalous dimension and to investigate in more 
details  the dependence of the amplitudes on the meson wave functions.
Also the real parts of the helicity amplitudes have to be considered.
    
\vspace*{1cm}

{\it \bf  Acknowledgements.}
In the final stage of this work we had encouraging discussions
with P. Marage, K. Piotrzkowski and J. Crittenden. 
D.Yu. I. would like to acknowledge the warm hospitality
extended to him at University of Leipzig.

\setcounter{equation}{0}

\end{document}